\documentclass[amsmath,amssymb,aps,prapplied,nonbibnotes,longbibliography,reprint,superscriptaddress]{revtex4-2}

\usepackage[english]{babel}
\usepackage[utf8]{inputenc}
\usepackage[T1]{fontenc}
\usepackage{graphicx}
\usepackage{xcolor}
\usepackage{soul}
\usepackage[separate-uncertainty=true]{siunitx}  %use \pm sign
\usepackage[version=4]{mhchem}
\usepackage[breaklinks=true,colorlinks=true,linkcolor=blue,urlcolor=blue,citecolor=blue]{hyperref}
\usepackage[normalem]{ulem} % fix problem with soul inside math
\newcommand{\equalcontribution}{These authors contributed equally to this work.}
\renewcommand{\selectlanguage}[1]{}  %resolves problem in babel with "language=en" or similar.

\begin{document}

\title{Tailoring the properties of YBa$_{2}$Cu$_{3}$O$_{7-\delta}$ thin films by 30-keV He$^+$ irradiation:\\ An enabling route to superconducting-device nanopatterning}

\author{Bernd Aichner}
\thanks{\equalcontribution}
\affiliation{Faculty of Physics, University of Vienna, 1090 Vienna, Austria}

\author{Simon Koch}
\thanks{\equalcontribution}
\affiliation{Physikalisches Institut, Center for Quantum Science (CQ) and LISA$^+$, Universität Tübingen, 72076 Tübingen, Germany}

\author{Philipp A. Korner}
\thanks{\equalcontribution}
\affiliation{Faculty of Physics, University of Vienna, 1090 Vienna, Austria}

\author{Max Karrer}
\affiliation{Physikalisches Institut, Center for Quantum Science (CQ) and LISA$^+$, Universität Tübingen, 72076 Tübingen, Germany}

\author{Katja Wurster}
\affiliation{Physikalisches Institut, Center for Quantum Science (CQ) and LISA$^+$, Universität Tübingen, 72076 Tübingen, Germany}

\author{Christoph Schmid}
\affiliation{Physikalisches Institut, Center for Quantum Science (CQ) and LISA$^+$, Universität Tübingen, 72076 Tübingen, Germany}

\author{Ulrich Kentsch}
\affiliation{Ion Beam Center (IBC), Institute of Ion Beam Physics and Materials Research, Helmholtz-Zentrum Dresden-Rossendorf (HZDR), 01328 Dresden, Germany}

\author{Reinhold Kleiner}
\affiliation{Physikalisches Institut, Center for Quantum Science (CQ) and LISA$^+$, Universität Tübingen, 72076 Tübingen, Germany}

\author{Edward Goldobin}
\affiliation{Physikalisches Institut, Center for Quantum Science (CQ) and LISA$^+$, Universität Tübingen, 72076 Tübingen, Germany}

\author{Dieter Koelle}
\affiliation{Physikalisches Institut, Center for Quantum Science (CQ) and LISA$^+$, Universität Tübingen, 72076 Tübingen, Germany}

\author{Wolfgang Lang} \email{wolfgang.lang@univie.ac.at}
\affiliation{Faculty of Physics, University of Vienna, 1090 Vienna, Austria}

{\onecolumngrid
This is an author-created version of an article published in Phys.~Rev.~Applied~{\bf 26},~044076~(2026). The Version of Record is available online free of charge under the \href{https://creativecommons.org/licenses/by/4.0/}{CC BY 4.0} license at \href{https://journals.aps.org/prapplied/pdf/10.1103/7dhm-9tvs}{https://journals.aps.org/prapplied/pdf/10.1103/7dhm-9tvs}. \hrule}
%\date{\today}

\begin{abstract}
Focused helium ion beam (He-FIB) irradiation with 30-keV ions is a key tool for nanoscale patterning and defect engineering in high-transition temperature ($T_\mathrm{c}$) cuprate superconducting devices, yet its usable fluence window is constrained by the competing requirements of reliable superconductivity suppression and minimal structural degradation. In this work, we provide a comprehensive dataset on the effects of large-area 30-keV He$^+$ ion exposure on the electric transport and superconducting properties of epitaxial YBa$_2$Cu$_3$O$_{7-\delta}$ (YBCO) thin films. X-ray diffraction shows a fluence-driven loss of crystalline order accompanied by an out-of-plane lattice expansion and an orthorhombic-to-tetragonal transition, culminating in predominant amorphization at the highest fluence of \qty{1e16}{\per\cm\squared}. Raman spectra exhibit increasing disorder while lacking signatures of oxygen depletion, indicating that irradiation mainly generates oxygen-related Frenkel defects rather than changing the carrier concentration. Consistently, with increasing fluence, the normal-state resistivity $\rho_\mathrm{N}(T)$ at temperature $T$ above $T_{\mathrm{c}}$ increases strongly, while $\partial\rho_\mathrm{N}/\partial T$ remains nearly unchanged at moderate fluence. The suppression of $T_{\mathrm{c}}$ is accurately described by Abrikosov--Gor'kov pair breaking and reaches complete quenching of superconductivity at $\qty{4.5e15}{\per\cm\squared}$. The anisotropic upper critical fields decrease approximately exponentially with increasing fluence, the vortex activation energy is reduced, and the anisotropy drops, in contrast to oxygen-depleted YBCO. Hall-angle analysis confirms a nearly constant carrier density but a systematic increase in defect scattering and reduced mobility, consistent with a crossover toward the dirty limit at high fluence. These results establish quantitative fluence thresholds and a practical operational window for He-FIB nanopatterning of YBCO quantum circuits.
\end{abstract}

\flushbottom
\maketitle

\section{Introduction}

Nearly four decades after the discovery of copper-oxide high-temperature superconductors \cite{BEDN86} and despite extensive experimental and theoretical investigation, the microscopic mechanism underlying high-temperature superconductivity (HTS) remains unresolved. Nevertheless, these materials---most notably the prototypical copper-oxide compound YBa$_{2}$Cu$_{3}$O$_{7-\delta}$ (YBCO)---are now thoroughly characterized and have already been implemented in technological applications. In particular, HTS electronic devices commonly rely on micro- and nanopatterned YBCO thin films as a fundamental platform \cite{KOEL99R,MART17R,DOBR26R}.

Focused helium ion beam (He-FIB) irradiation has emerged as a highly versatile method for tailoring the superconducting properties of YBCO thin films with nanometer-scale spatial precision, thereby enabling device architectures that are challenging to implement using conventional subtractive patterning techniques. A principal advantage of light-ion (He$^+$) irradiation processing is the ability to systematically tune superconducting properties via irradiation-induced point defects \cite{ARIA03,LANG06a}, predominantly oxygen Frenkel pairs \cite{GRAY22}, while maintaining both the film surface and the underlying crystallographic framework essentially unaltered at moderate ion fluence \cite{MULL19, SCHM26P, ZALU24}. This contrasts with conventional lithographic nanopatterning approaches, which inherently involve thickness inhomogeneities and the creation of exposed interfaces that can promote oxygen depletion and thereby deteriorate the electronic transport characteristics.

In state-of-the-art helium ion microscopes (HIMs) \cite{WARD06,HLAW16M}, the He-FIB enables the direct fabrication of spatially alternating superconducting and insulating domains, with lateral feature sizes on the order of only a few nanometers. Exploiting this capability, single Josephson junctions (JJs) \cite{CYBA15,MULL19,KARR24,SCHM25}, Josephson diodes \cite{SCHM25}, JJ arrays \cite{LEFE19,PROP25}, superconducting quantum interference devices (SQUIDs) \cite{CHO15,MULL19,SCHM26P}, and directional in-plane tunnel spectroscopy \cite{LEFE23} have been realized. Furthermore, He-FIB–engineered artificial vortex pinning landscapes in YBCO \cite{AICH19} have recently been shown to sustain pronounced vortex pinning effects up to magnetic fields of \qty{6}{T} \cite{KARR24a}, while also exhibiting robust temporal stability over time scales of several years \cite{KEPP24}.

To date, He-FIB nanopatterning has predominantly relied on empirically guided selection of irradiation parameters. However, achieving device-to-device reproducibility necessitates a detailed understanding of the concomitant structural evolution, the superconducting and normal-state transport characteristics, and their correlation with the nature and density of irradiation-induced defects. In He-FIB processing using a HIM, an ion energy of 30\,keV is typically employed. Although several studies have examined the influence of higher-energy He$^+$ irradiation on the properties of thin YBCO films \cite{XIAO96a,ARIA03,SEFR01,LANG06a,ZALU24}, a comprehensive analysis of the effects of 30-keV He$^+$ irradiation is still lacking.

Our study aims to establish a robust foundation for subsequent applications of the He-FIB technique by providing a comprehensive dataset on YBCO thin films irradiated with 30-keV He$^+$ ions over large areas. We systematically investigate the modifications of the $a$, $b$, and $c$ crystallographic lattice parameters, the evolution of Raman-active modes and spectral background as a function of ion fluence, and changes in the electrical transport properties through magnetoresistivity and Hall-effect measurements, including derived estimates of the anisotropy and upper critical fields. Taken together, these results provide quantitative guidelines for optimizing irradiation conditions in the current and future fabrication of functional nanostructures using a HIM.

%---------------------------------------------------
\section{Methods}

Thin YBCO films were epitaxially grown on (LaAlO$_3$)$_{0.3}$(Sr$_2$AlTaO$_6$)$_{0.7}$ (LSAT) substrates by pulsed laser deposition (PLD). X-ray diffraction (XRD) $\theta-2\theta$ scans yield $c$-axis oriented films, and Laue oscillations at the YBCO (001) Bragg peak were used to determine the film thickness $t$. The full width at half maximum of typically $\lesssim \ang{0.1}$ of the rocking curve from the YBCO (005) peak \cite{KARR24a} confirmed excellent epitaxial growth of the films.

For XRD analysis of structural modifications induced by He-ion irradiation, we grew two 50-nm-thick YBCO films on \qtyproduct{10 x 10}{\mm} LSAT substrates and cut each of them into six \qtyproduct{4.5 x 3}{\mm} small chips. Two of them (one from each batch) were examined by XRD prior to irradiating all 12 small chips at the Ion-Beam Center of the Helmholtz-Zentrum Dresden-Rossendorf, Germany. 

Ion irradiation at room temperature was carried out using an ion implanter (Danfysik A/S, Denmark, Model 1050) with He$^+$ ions at an energy of 30-keV and ion fluences $\Phi$ in the range from \qty{5e14}{\cm^{-2}} to \qty{1e16}{\cm^{-2}} (\qtyrange{5}{100}{\per\nm\squared}), incident orthogonally to the film surface. The fluence was monitored by Faraday cups with an accuracy of 4\%.  Stopping and Range of Ions in Matter and Transport of Ions in Matter (SRIM/TRIM) simulations \cite{ZIEG10} were performed with atomic displacement energies of \qty{35}{\eV} for yttrium and barium (Ba) \cite{KIRS92}, \qty{15}{\eV} for copper (Cu)(1), and Cu(2) \cite{LEGR93}, \qty{8.4}{\eV} for oxygen (O)(2), O(3), and O(4) \cite{TOLP96}; and \qty{1}{\eV} for O(1) \cite{ROTH89}. The results indicate that essentially all 30-keV He$^+$ ions traverse YBCO films with thickness $t \lesssim \qty{50}{\nm}$ and are only decelerated and implanted within the LSAT substrate.

Following ion irradiation, the samples were stored for more than 2 weeks in a dry nitrogen atmosphere at ambient temperature to facilitate the annealing of defects with low activation energies. This protocol is motivated by our previous observation that a fraction of the ion-induced atomic displacements recombines at room temperature on a timescale of days, whereas the remaining defects persist over timescales of years \cite{KARR24,KEPP24}. All XRD measurements on irradiated samples were conducted within 3 weeks in order to minimize residual relaxation processes occurring between successive data acquisitions.

The crystal structure of the YBCO thin films was examined by XRD using an X'Pert MRD Pro diffractometer (Philips), equipped with a PIXel1D line detector (Malvern Panalytical GmbH), and employing Cu K$\alpha_1$ radiation with a wavelength of $\lambda = 1.540598$\,\AA. All (00$l$) peaks with sufficient intensity were used for subsequent analysis of the $c$-axis lattice parameter. Owing to a slight orthorhombic distortion in the $a$ and $b$ lattice parameters, the YBCO films are twinned, i.e., the $a$ and $b$ crystallographic directions are interchanged across domains. Consequently, separate (308) and (038) Bragg peaks were observed in the $2\theta$--$\omega$ two-axis scans. For lower ion fluences $\Phi < \qty{5e15}{\cm^{-2}}$, the diffracted intensities were first integrated over the $\omega$ angle, and the resulting intensity-versus-$2\theta$ profiles were smoothed using a Savitzky–Golay filter to identify and analyze the position of the two Bragg peaks. At higher fluences, the (308) and (038) peaks progressively merged and therefore had to be analyzed manually.

For the electrical characterization, two YBCO thin films with $t_\mathrm{A} = \qty{28}{\nm}$ (chip A) and $t_\mathrm{B} = \qty{30}{\nm}$ (chip B) were grown on \qtyproduct{10 x 10}{\mm} LSAT substrates and capped \emph{in-situ} with 20-nm-thick gold (Au) layers deposited by electron beam evaporation immediately after PLD growth. In a first patterning step, by photolithography and argon (Ar) ion beam etching, we milled the Au and YBCO bilayers on chip A to produce 12 bridge structures of \qty{8}{\um} width and \qty{40}{\um} length, with voltage probes separated by \qty{20}{\um}. On chip B, four bridge structures with dimensions \qtyproduct{40 x 180}{\um} and voltage probes separated by \qty{70}{\um} were patterned. Subsequently, in a second lithography step, the Au capping layer was removed from the microbridge regions using Lugol’s iodine solution, thereby providing direct access to the underlying YBCO film for irradiation while preserving the integrity of the electrical contact regions.

The YBCO microbridges on chip A were subsequently processed using a Zeiss Orion NanoFab HIM. The 30-keV He$^+$ ion beam was defocused by moving the beam focus point by +\qty{25}{\um}, thereby increasing its diameter on the sample surface to approximately \qty{350}{\nm}. This setting was used to irradiate a square lattice with a point-to-point spacing of \qty{12}{\nm} over an area of \qtyproduct{100 x 40}{\um}. The irradiated region fully encompassed the width of the bridge and extended beyond the locations of the voltage probes. Under these conditions, spatial variations in the local ion fluence are negligible, such that the effective areal ion fluence conforms to broad-beam irradiation and is governed solely by the total number of ions incident on the bridge. Consequently, the irradiation fluence values achieved with the ion implanter and the HIM are equivalent. The temporal stability of the gas field ion source current in the HIM is approximately 2\% \cite{KLIN20}, which determines the accuracy of the delivered fluence. Chip B was used for reference measurements of an unirradiated bridge.

Magnetoresistivity and Hall-effect measurements were carried out using a Physical Property Measurement System (Quantum Design) equipped with a \qty{9}{\tesla} superconducting solenoid and a variable-temperature insert. Electrical contacts to the sample holder were prepared with 50-$\mu$m-diameter Au wires and silver paste, and the holder was mounted on a horizontal rotator. The angular scale of the rotator was calibrated by monitoring the pronounced minimum in the magnetoresistance associated with the intrinsic vortex lock-in transition \cite{TACH89}, which occurs when the applied magnetic field is precisely aligned parallel to the CuO$_2$ planes. To suppress spurious thermoelectric contributions, all measurements were performed for both current polarities. Hall measurements were additionally conducted for both polarities of the magnetic field, and the Hall voltage was obtained as half of the difference between the corresponding transverse voltage signals.

Raman spectra were acquired at room temperature on the same specimens used for the transport measurements employing a WITec alpha 300A system, which integrates a Raman spectrometer with a high‑resolution confocal optical microscope. A \qty{532}{\nm} excitation laser with a power of \qty{3}{\mW} was coupled to the sample surface via a single-mode optical fiber and focused onto the specimen by the microscope optics. The backscattered radiation was collected through the same objective lens, coupled into a multimode optical fiber, and guided to a spectrometer equipped with a \qty{600}{grooves/\mm} diffraction grating and a CCD detector. For each sample, Raman spectra were recorded at three distinct positions and subsequently averaged. Owing to the small thickness of the YBCO film, a substantial background contribution originating from the substrate was observed and numerically subtracted. The signal-to-noise ratio was optimized by employing the maximum laser power that did not induce any detectable modification or degradation of the sample surface, as verified by the reproducibility of successive spectra acquired at the same location.

%---------------------------------------------------
\section{Results and Discussion}

\begin{figure}[h!]
\centering
\includegraphics[width=0.6\columnwidth]{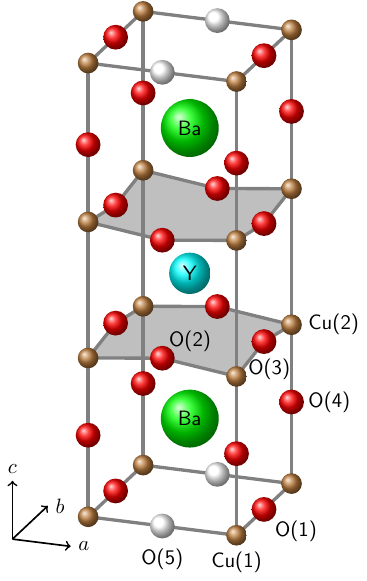}
\caption{Unit cell of YBCO in the orthorhombic phase. The CuO$_2$ planes, highlighted in gray, host superconductivity, enabled by hole-charge transfer from the Cu(1)-O(1) chains in the basal planes. The vacant O(5) positions are shown in white.}
\label{fig:YBCO}
\end{figure}

The set of samples examined in this study spans unirradiated, pristine YBCO thin films to specimens exposed to ion irradiation with a fluence of $\Phi = \qty{1e16}{\cm^{-2}}$, a value at which superconductivity is completely suppressed and the YBCO crystal structure is, to a large extent, destroyed. The YBCO unit cell and the conventional notation for the atomic sites are illustrated in Fig.~\ref{fig:YBCO}.

Two fluence values are essential for the design of superconducting devices: on the one hand, the complete suppression of superconductivity enables the creation of barriers in Josephson junctions or pinning sites for Abrikosov vortices. On the other hand, amorphization of the crystalline structure is unfavorable, as amorphous tracks created by He-FIB grow in diameter with increasing numbers of impacting ions \cite{MULL19} and deteriorate the method's resolution \cite{SCHM26P}. Thus, the operational range is typically selected between these two fluence values.  

%---------------------------------------------------
\subsection{X-ray diffraction}
\label{sec:XRD}

\begin{figure*}[ht!]
\centering
\includegraphics[width=\linewidth]{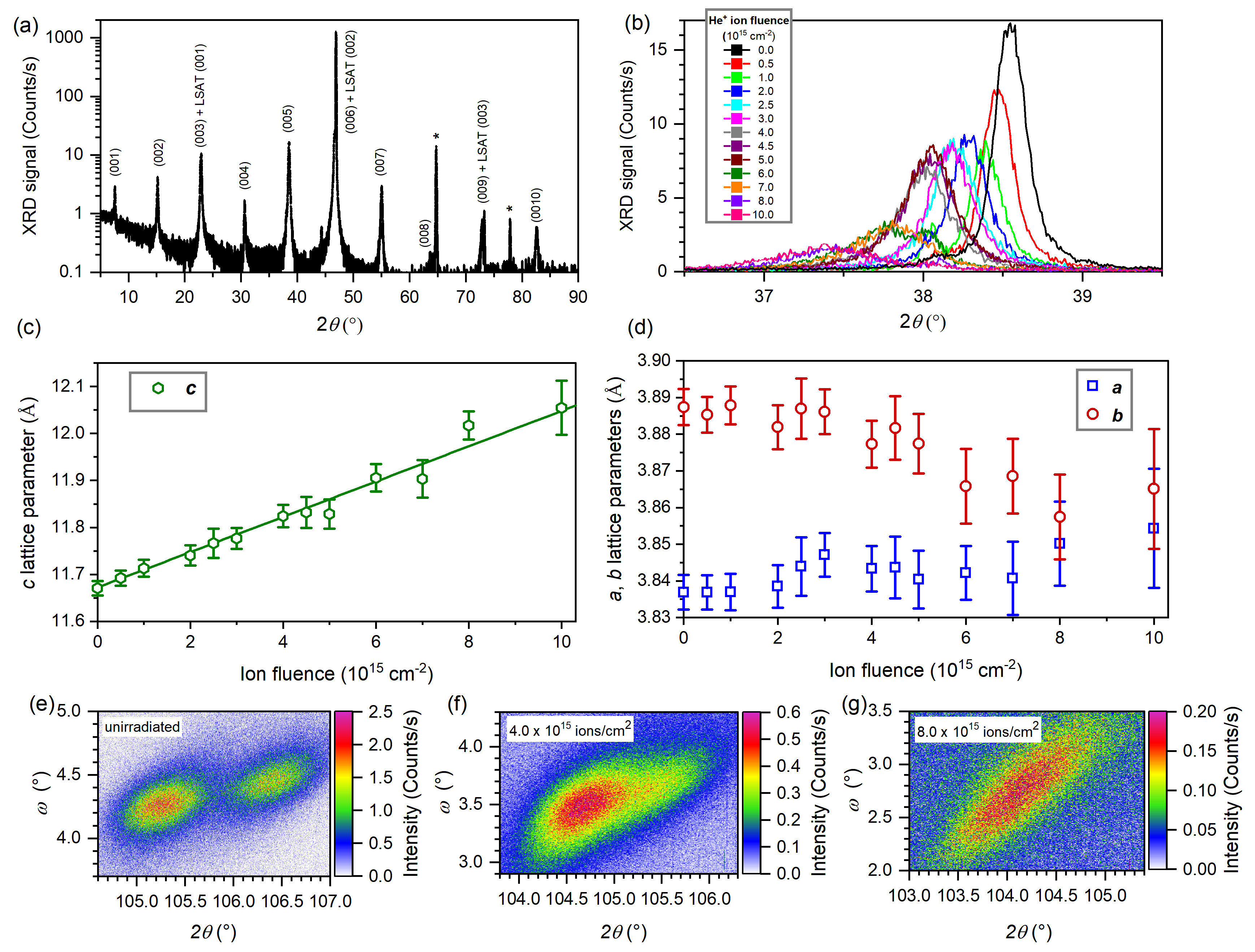}
\caption{XRD analysis of YBCO thin films deposited on LSAT substrates. (a) $(00l)$ reflection peaks in the $\theta$–$2\theta$ scan of a pristine reference film. The asterisks denote peaks that are presumably associated with the aluminum sample holder. (b) Evolution of the (005) peak upon irradiation with 30-keV He$^+$ ions at various fluences. (c) Evolution of the out-of-plane lattice parameter $c$ inferred from the positions of the (00$l$) peaks. The line represents a linear regression. (d) Variation of the in-plane lattice parameters $a$ and $b$ of YBCO as a function of fluence, determined from two-axis scans of the (308) and (038) peaks. (e)-(g) Two-axis ($2\theta$–$\omega$) maps of the (308) and (038) diffraction peaks for the pristine film and films irradiated with fluences of \qty{4.0e15}{\cm^{-2}} and \qty{8.0e15}{\cm^{-2}}, respectively.}
\label{fig:XRD}
\end{figure*}

The high crystalline quality and epitaxial growth of the thin YBCO films are evidenced by the sharp $(00l)$ diffraction peaks shown in the $\theta$–$2\theta$ scan of a pristine reference film in Fig.~\ref{fig:XRD}(a). Upon He$^+$ irradiation, the intensities of the $(00l)$ peaks decrease, their widths increase, and their positions shift to lower diffraction angles, as displayed in Fig.~\ref{fig:XRD}(b) for the YBCO (005) peak. In particular, for ion fluences $\Phi > \qty{5.0e15}{\cm^{-2}}$, the intensity of the (005) peak is significantly reduced and is almost fully suppressed for $\Phi > \qty{7.0e15}{\cm^{-2}}$. From the diffraction peak positions, the out-of-plane $c$-axis lattice parameter was determined and is observed to increase approximately linearly as a function of ion fluence, as shown in Fig.~\ref{fig:XRD}(c).

A comparable, nearly linear expansion of the lattice is well established from investigations of oxygen-deficient YBCO \cite{JORG90a,YE93}, where the reduced occupancy of the O(1) sites diminishes the strength of the covalent bonding within the stacked YBCO structure. There is, however, a notable distinction: whereas the $c$-axis lattice parameter increases from 11.67\,\AA\ to 11.83\,\AA\ with increasing oxygen deficiency in the range $0.05 < \delta < 0.95$, He$^+$ irradiation leads to an expansion to significantly larger $c \sim (12.05 \pm 0.06$)\,\AA\ at $\Phi = \qty{1.0e16}{\cm^{-2}}$. Although the latter value must be interpreted with caution due to the predominant amorphization of the YBCO film under these conditions, the more pronounced unit-cell expansion clearly reflects the distinct microscopic mechanisms underlying oxygen depletion versus ion irradiation. As discussed in detail in Secs. \ref{sec:resist} and \ref{sec:Hall}, the electrical transport measurements indicate that irradiation-induced oxygen loss is negligible, and instead point to enhanced structural disorder as the primary origin of the observed lattice expansion.

The XRD measurements also enable the determination of the in-plane lattice parameters $a$ and $b$, which are presented in Fig.~\ref{fig:XRD}(d). While the crystal structure along the $a$ axis exhibits only a marginal expansion with increasing ion fluence, the $b$ lattice parameter decreases, resulting in an orthorhombic-to-tetragonal (O-T) phase transition at approximately $\Phi \approx \qty{8.0e15}{\cm^{-2}}$. This structural evolution is illustrated in Figs.~\ref{fig:XRD}(e)–\ref{fig:XRD}(g) by false-color maps of the two-axis ($2\theta-\omega$) diffraction intensities. In the pristine film, two well-separated diffraction maxima are observed; these begin to partially merge at $\Phi = \qty{4.0e15}{\cm^{-2}}$ and fully coalesce at $\Phi = \qty{8.0e15}{\cm^{-2}}$. Upon irradiation, both peaks shift to smaller $2\theta$ angles, consistent with an increase in the out-of-plane $c$-axis lattice parameter.

The comparatively low XRD signal intensities render an unambiguous determination of the O-T phase transition challenging. Beyond this instrumental limitation, a broadening of the diffraction peaks is expected, which can be attributed to local variations in the orthorhombicity of the unit cell. Molecular dynamics simulations of irradiation-induced damage \cite{GRAY22} suggest that Frenkel defects are generated by displacing CuO chain oxygen atoms from the O(1) site into the initially vacant O(5) positions within the same crystallographic plane of the unit cell. At moderate ion fluences, this process may not be fully stochastic, thereby promoting the formation of finite CuO chain segments. These segments, in turn, lead to an anisotropic modification of the lattice, giving rise to a disproportionate change in the $a$ and $b$ lattice parameters. The rearrangement of such CuO chain segments has been proposed as one of the mechanisms underlying persistent photoconductivity in oxygen-deficient YBCO \cite{MARK96}, a phenomenon that has indeed also been observed in YBCO irradiated with 75-keV \cite{MARK10} and 80-keV \cite{NAVA00} He$^+$ ions.

Notably, an O-T phase transition is not necessarily associated with oxygen depletion \cite{WANG95b}. The distinction in the present case is that, under ion irradiation, the in-plane lattice parameter $a \simeq 3.84$\,\AA\ remains essentially invariant, whereas the parameter $b$ exhibits an approximately linear decrease as a function of increasing fluence. Taking into account the error bars in Fig.~\ref{fig:XRD}(d), together with the merging of the (308) and (038) diffraction peaks in the $2\theta$--$\omega$ scans shown in Fig.~\ref{fig:XRD}(g), we infer that the films undergo a transition to a tetragonal structure for $\Phi \geq \qty{8.0e15}{\cm^{-2}}$.  

In oxygen-deficient YBCO, by contrast, the lattice parameter $a$ increases while $b$ decreases, and the two become equal only for $\delta \gtrsim 0.7$. An XRD investigation of YBCO films irradiated with 90-keV He$^+$ ions reported essentially no variation in $a$ and $b$ for $\Phi \leq \qty{5.0e15}{\cm^{-2}}$, followed by an abrupt orthorhombic-to-tetragonal transition with $a=b$ for $\Phi \geq \qty{5.0e15}{\cm^{-2}}$ \cite{ZALU24}.

Moreover, recent nanofocused XRD measurements \cite{ZALU24,CART26} have revealed that variations in the lattice parameters also depend on the lateral dimensions of the irradiated region. This behavior has been attributed to the elastic strain energy arising at the interface between the irradiated and non-irradiated (pristine) areas of the film, which can induce crystal-lattice distortions that interact with ion-irradiation-induced defects. Indeed, recent findings indicate that strain effects play a significant role in governing the magnetic and electric properties of YBCO \cite{WAHL26}.

%---------------------------------------------------
\subsection{Raman scattering}
\label{sec:Raman}

\begin{figure*}[ht]
\centering
\includegraphics[width=0.8\linewidth]{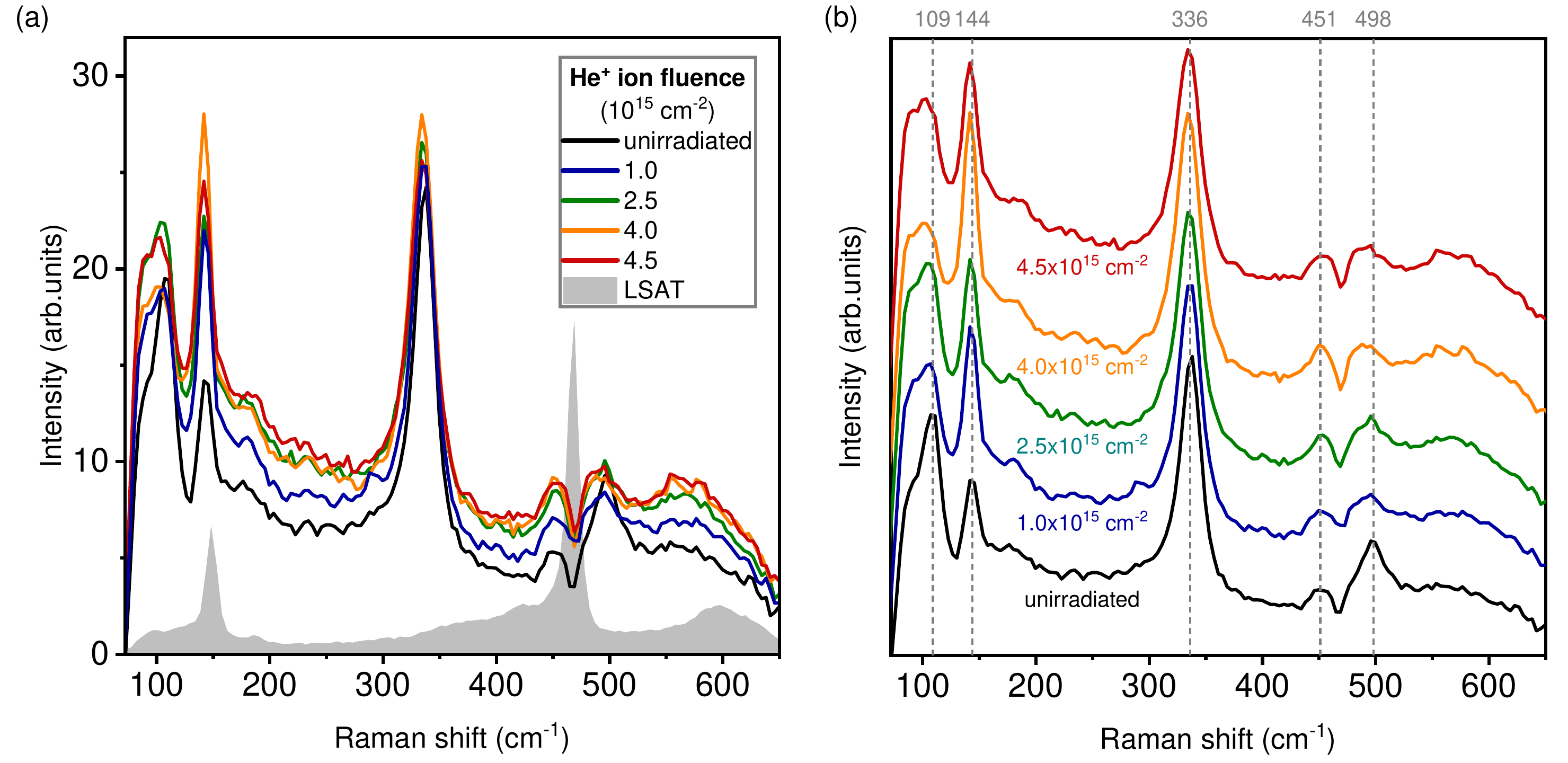}
\caption{(a) Raman spectra of YBCO bridges subjected to different irradiation conditions. The gray region represents the Raman response of the LSAT substrate, acquired from the same sample (intensity not to scale). (b) The five Raman spectra are displayed with uniform vertical offsets for clarity. The dashed lines indicate Raman-active modes discussed in the main text. Each spectrum corresponds to the average of three measurements acquired at different positions on the same bridge. A baseline correction has been applied, and the substrate contribution has been subtracted.}
\label{fig:Raman}
\end{figure*}

Raman spectroscopy is an effective method for probing local structural modifications induced by irradiating YBCO with 30-keV He$^{+}$ ions. However, it must be noted that the probing laser also penetrates into the underlying substrate, whose Raman-active modes contribute to the overall measured signal. LSAT is regarded as a suitable substrate for Raman investigations, as it exhibits comparatively broad spectral windows devoid of phonon modes, combined with a weak and spectrally flat Raman background \cite{GASP14}. Reference measurements performed on an uncovered region of the substrate identified Raman modes of LSAT at \qty{148}{\per\cm} and \qty{469}{\per\cm}, as indicated by the gray-shaded area in Fig.~\ref{fig:Raman}(a).

Figure~\ref{fig:Raman} presents the Raman spectra of a pristine reference sample and multiple YBCO bridges fabricated on the same chip A and subsequently irradiated with 30-keV He$^{+}$ ions at different fluences. In all spectra, the contribution of the substrate, obtained from an independent reference measurement, as well as the baseline of the Raman signal, has been subtracted. Note that the samples are the same as those previously employed for the transport measurements discussed in Secs.~\ref{sec:resist}–\ref{sec:Hall}.

In the Raman spectrum of the unirradiated sample, as shown in Figs.~\ref{fig:Raman}(a) and \ref{fig:Raman}(b), five vibrational modes are discernible. The line at \qty{109}{\per\cm} originates from vibrations of the Ba atoms \cite{LIAR00}, whereas the feature at \qty{144}{\per\cm} is attributed to the Cu(2)-A$_g$ vibrational mode \cite{HONG10}. The latter is a well-established probe of the oxygen content and exhibits a shift toward lower wave numbers upon oxygen depletion. Its spectral weight increases with the volume fraction of the tetragonal phase. Although the peak position changes only slightly from \qty{144}{\per\cm} to \qty{142}{\per\cm} at the highest fluence, its spectral weight increases significantly, indicating that the oxygen stoichiometry remains essentially unchanged while the fraction of tetragonal unit cells in the samples grows. Oxygen-deficient YBCO is characterized by a Raman mode at \qty{228}{\per\cm}, arising from vibrations of oxygen copper atoms located at the terminations of disrupted Cu–O chains \cite{CHRO17}; this feature is absent in our data, and we hypothesize that irradiation-induced redistribution of oxygen from the O(1) to the O(5) sites may suppress this Cu vibrational mode, in contrast to oxygen-deficient YBCO, in which the O(5) sites remain essentially unoccupied. Even considering a possible contribution of the LSAT substrate peak at \qty{148}{\per\cm} that could influence the detailed line shape of the \qty{144}{\per\cm} phonon, the lack of any detectable \qty{228}{\per\cm} signal provides strong evidence against significant oxygen depletion during irradiation.

The vibrational mode at \qty{336}{\per\cm} is associated with out-of-phase displacements of the O(2) and O(3) atoms. A decrease in oxygen content is expected to manifest as a transformation of the corresponding spectral line shape from a Fano profile to a symmetric Lorentzian function \cite{ILIE97}. Although such a transition is not detected in the present measurements, a slight shift of the O(2,3)-A$_g$ phonon from \qty{336}{\per\cm} to approximately \qty{334}{\per\cm} is observed. This behavior, albeit infrequently reported, has been ascribed to tensile stress arising from microstructural damage induced by ion irradiation \cite{WANG19}.

Resonances at approximately \qty{451}{\per\cm} and \qty{498}{\per\cm} can be assigned to the tetragonal and orthorhombic phases of YBCO, respectively \cite{HONG10}. While the former may be perturbed by the proximity of the LSAT substrate line, the latter is associated with vibrational modes of the apical O(4) atoms along the crystallographic $c$ axis. These apical oxygens constitute the bridging link between the CuO chains and the CuO$_{2}$ planes. The spectral weight of this peak decreases markedly already at a fluence of $\Phi = \qty{1e15}{\cm^2}$. At higher fluence, the apparent broadening of the feature is more consistently interpreted as arising from the emergence of a second resonance at \qty{484}{\per\cm}. Analogous peak splitting has previously been attributed to the Ortho-II phase in oxygen-deficient YBCO \cite{ILIE97}, which is characterized by alternating fully occupied and empty CuO chains. However, since our measurements provide no evidence for substantial oxygen depletion under any of the investigated conditions, we instead propose that this behavior originates from stochastic migration of oxygen atoms from O(1) to O(5) sites.

Two additional features in the Raman spectrum [not marked in Fig.~\ref{fig:Raman}(b)] merit attention. The first is an emergent peak at \qty{92}{\per\cm}, which has not been reported in the experimental literature and may be associated with low-energy vibrational modes of the Ba atom parallel to the $ab$ plane, according to theoretical calculations \cite{LIU88,BATE89}. The second concerns a broad band centered at approximately \qty{560}{\per\cm}, for which a consensus on the underlying phonon modes has not yet been reached. However, its assignment to the formation of oxygen vacancies at the O(1) sites, as inferred from oxygen-depletion experiments \cite{MCCA88}, is further corroborated by our data, which show an enhancement of this band upon irradiation.

Finally, the enhancement of the background intensity evident in Fig.~\ref{fig:Raman}(a) represents a generic spectroscopic hallmark of irradiation-induced increase in structural disorder. Taken together, the Raman scattering data substantiate the interpretation that the interaction of incident He$^+$ ions with the YBCO crystal lattice leads predominantly to displacements of O(1) atoms, presumably accompanied by partial occupation of O(5) sites. The overall crystallographic framework remains largely intact, at least for irradiation fluences $\Phi \leq \qty{4.5e15}{\per\cm\squared}$, as evidenced by the negligible changes in the \qty{336}{\per\cm} mode associated with vibrations of atoms in the CuO$_2$ planes.

%---------------------------------------------------
\subsection{Resistivity}
\label{sec:resist}

\begin{figure*}[t!]
\centering
\includegraphics[width=\linewidth]{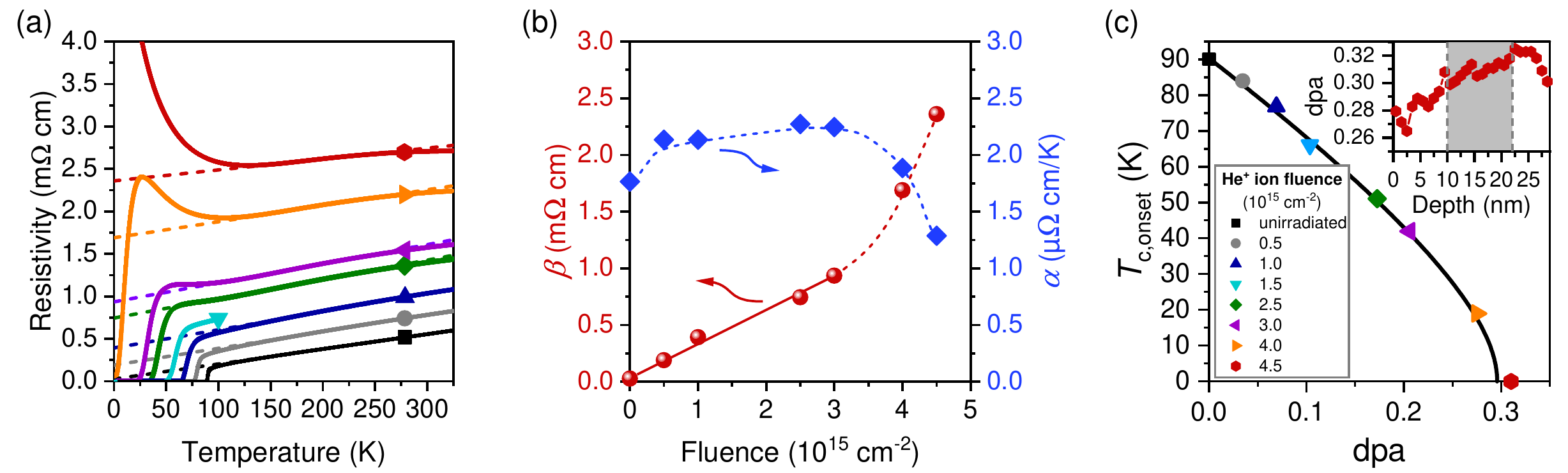}
\caption{Evolution of electric transport properties of YBCO microbridges with ion fluence [colors and symbols are defined in panel (c)]. (a) $\rho(T)$ curves. Broken lines are linear fits to the normal-state resistivity $\rho_\mathrm{N}(T)$. (b) Intercepts $\beta$ and slopes $\alpha$ obtained from the linear fits in panel (a). The solid line represents a linear fit, and the broken lines are guides to the eye. (c) $T_{\mathrm{c,onset}}$ vs displacements per atom (dpa). The solid line represents a fit based on the Abrikosov–Gor'kov pair-breaking theory, Eq.~\eqref{eq:AG}. Inset: Representative depth profile of the dpa extracted from simulations. The data within the depth interval indicated by the gray-shaded region are averaged to obtain the effective dpa values displayed in the main panel.}
\label{fig:resistance}
\end{figure*}

Modifying the electrical transport properties of YBCO in both its superconducting and normal states constitutes a primary objective of ion-irradiation studies. Figure \ref{fig:resistance}(a) displays the evolution of the resistivity–temperature dependence $\rho(T)$ before and after irradiation with 30-keV He$^{+}$ ions at various fluences. Irradiation-induced defects give rise to an enhancement of the normal-state resistivity $\rho_\mathrm{N}(T)$ accompanied by a suppression of the critical temperature $T_\mathrm{c}$. At low fluences, the slope of $\rho_\mathrm{N}(T)$ remains nearly unchanged; however, for fluences exceeding \qty{3.0e15}{\per\cm\squared}, it decreases, and the initially metallic temperature dependence progressively evolves into a thermally activated hopping behavior at low temperatures \cite{TOLP96a}. Remarkably, at a fluence of \qty{4.5e15}{\per\cm\squared}, $\rho_\mathrm{N}(T)$ becomes nearly temperature independent in the range $\qty{120}{K} < T < \qty{325}{K}$, a characteristic that may be advantageous for specific device applications.

In accordance with prior studies, the normal-state resistivity of superconducting YBCO (as well as that of most other HTS) can, to a very good approximation, be represented by:
\begin{equation}
\label{eq:rho}
  \rho_\mathrm{N}(T)=\alpha T + \beta,  
\end{equation}
where the slope $\alpha$ is inversely proportional to the normal-state carrier concentration \cite{ANDO04}, whereas the intercept $\beta$ represents a temperature-independent carrier scattering rate \cite{CHIE91a}. The value $\beta = \qty{27.6}{\micro\ohm\cm}$ obtained for the pristine sample is similar to reports on 50-nm-thin YBCO films \cite{SEFR01} and indicates a low density of carrier-scattering defects. Upon irradiation, $\beta$ increases linearly with fluence for $\Phi \leq \qty{3.0e15}{\per\cm\squared}$, as shown in Fig.~\ref{fig:resistance}(b). The subsequent upturn of $\beta(\Phi)$ at higher fluence is most likely associated with an increasing prevalence of the thermally activated $\rho(T)$ behavior, which renders the simple linear relation in Eq.~\eqref{eq:rho} inapplicable in this regime.

The negligible variation of the parameter $\alpha$ up to $\Phi \le \qty{3e15}{\per\cm\squared}$ indicates that the charge carrier concentration remains essentially unchanged. Furthermore, the density of mobile carriers is directly connected to the oxygen content of YBCO \cite{ITO93}, corroborating the interpretation that the suppression of $T_\mathrm{c}$ is governed primarily by irradiation-induced defect formation, rather than by oxygen depletion. Moreover, the characteristic downturn (the so‑called ``S‑shape'' feature) in the $\rho_\mathrm{N}(T)$ curves of underdoped YBCO \cite{ITO93,ANDO04}, typically associated with the onset of the pseudogap \cite{SOLO06b}, is absent in the present data set. 

No evidence for multiple domains with different superconducting properties---typically expected to manifest as a shoulder in the resistive transition---is observed. This absence implies that the irradiation predominantly creates randomly distributed point defects that do not aggregate into extended defect clusters. By contrast, irradiation with heavier ions, such as \qty{1}{\MeV} Ne$^+$ ions, has been reported to induce a two-step superconducting transition, which has been attributed to an inhomogeneous spatial distribution of irradiation-induced defects \cite{CHEN89}. In the present samples, however, the superconducting transition, while remaining single-step, exhibits a noticeable broadening upon irradiation. This broadening is most plausibly ascribed to the strong sensitivity of the superconducting order parameter to local variations in defect density, a sensitivity that is amplified by the very short in-plane Ginzburg–Landau coherence length at $T=0$, $\xi_{ab}(0) < \qty{2}{\nm}$ \cite{SEKI95}, extending over only a few unit cells.

For a more detailed assessment of the impact of irradiation, we define a superconducting onset temperature, $T_{\mathrm{c,onset}}$, as the intersection between a fit to the normal-state resistivity, $\rho_\mathrm{N}(T)$, and a linear fit to the steepest segment of the superconducting transition. The resistivity at this intersection point is denoted as $\rho_\mathrm{onset}$ and is employed as an approximate parameter to model an abrupt superconducting transition commencing below $T_{\mathrm{c,onset}}$. Through this procedure, we aim to minimize the influence of transition rounding arising from fluctuations of the superconducting order parameter. The fitting procedure is described in detail elsewhere \cite{MLET19}. The evolution of $T_{\mathrm{c,onset}}$ as a function of dpa, obtained from SRIM and TRIM simulations, is shown in Fig.~\ref{fig:resistance}(c). At low fluences, $T_{\mathrm{c,onset}}$ exhibits an approximately linear decrease with increasing dpa; however, for fluences exceeding \qty{3.0e15}{\per\cm\squared}, it is more strongly suppressed and tends toward zero. At a fluence of \qty{4.5e15}{\per\cm\squared}, no signatures of superconductivity are observed down to \qty{2}{\K}.

The reduction of $T_{\mathrm{c,onset}}$ with increasing fluence is quantitatively described by the Abrikosov–Gor'kov pair-breaking theory \cite{ABRI60}. Although this theory was originally developed for magnetic impurities, it was subsequently predicted to be applicable as well to non-magnetic impurities in superconductors with $d$-wave order-parameter symmetry \cite{LARK65}. It is now firmly established that this theoretical framework also provides an adequate description of pair-breaking phenomena in HTS \cite{LESU90,TOLP96,TOLP96a}. The normalized critical temperature $T_{\mathrm{c,onset}}/T_0$, where $T_0=\qty{90.1}{\K}$ denotes the critical temperature of the pristine (unirradiated) superconducting film, can be represented as a function of the pair-breaking scattering time $\tau_\mathrm{p}$ \cite{TOLP96}:
\begin{equation}
\label{eq:AG}
\ln\Big(\frac{T_\mathrm{c,onset}}{T_0}\Big)=\Psi\Big(\frac{1}{2}\Big) - \Psi\Big(\frac{1}{2}+\frac{C}{4 \pi \tau_\mathrm{p} T_{\mathrm{c,onset}}} \Big),
\end{equation}
where $\Psi$ denotes the digamma function, $C$ is a proportionality constant with units of K\,s, and $\tau_\mathrm{p}$ is inversely proportional to the defect density, which in turn is correlated with the dpa value. It is significant to note that the fit to Eq.~(\ref{eq:AG}), represented by the solid line in Fig.~\ref{fig:resistance}(c), involves only a single adjustable parameter, namely the ratio $C/\tau_\mathrm{p}$.

The experimental data and corresponding fit presented in Fig.~\ref{fig:resistance}(c) can be utilized to quantitatively correlate SRIM and TRIM simulations with the measured $T_{\mathrm{c,onset}}$, as outlined in Ref.~\cite{MLET19}. Briefly, the simulations yield a depth-resolved damage profile in terms of dpa, which exhibits moderate variations due to the development of secondary collision cascades and a non-negligible contribution from the underlying substrate, as depicted in the inset of Fig.~\ref{fig:resistance}(c). Moreover, in YBCO thin films, some layers at the surface and at the interface to the substrate contribute only marginally to electric transport \cite{PUIC04a}. To reduce the influence of these effects, we calculated an average dpa value over the film thickness interval from \qty{10}{\nm} to \qty{22}{\nm}, and this averaged quantity was then compared with the experimentally determined $T_\mathrm{c,onset}$ at the corresponding ion fluence. Note that the experimental fluence and the dpa determined from simulations in Fig.~\ref{fig:resistance}(c) are related by $\Phi = \qty{1.453e16}{\per\cm\squared} \times \mathrm{dpa}$. 

The $T_\mathrm{c,onset}(\Phi)$ values are listed in Table~\ref{tab:properties} and establish a quantitative design parameter for the controlled modification of superconducting circuits via HIM irradiation of YBCO thin films. For ion fluences up to $\Phi \leq \qty{3.0e15}{\per\cm\squared}$, the superconducting transition temperature decreases linearly with increasing irradiation fluence, with a suppression rate of $\mathrm{d} T_\mathrm{c,onset}/\mathrm{d} \Phi \approx \qty{-1.68e-14}{\K\cm\squared}$.

Alternatively, we place our present results for 30-keV He$^+$ ion irradiation at normal incidence in the broader framework of nonionizing energy loss (NIEL), which quantifies the primary knock-on displacements produced by incident particles in the target lattice. A universal linear relationship spanning approximately 7 orders of magnitude in NIEL, and incorporating electron, proton, and heavy-ion irradiation, has been established to predict $\mathrm{d} T_\mathrm{c}/\mathrm{d}\Phi$ at moderate fluences for arbitrary particle species, kinetic energies, and fluences based solely on a NIEL calculation \cite{SUMM89}. Employing the same displacement threshold energies as in the TRIM and SRIM simulations, we obtain a NIEL value of \qty{10.4}{\MeV\cm\squared\per\g} using the SR-NIEL simulation package \cite{SR-NIEL}. On this basis, our measured $\mathrm{d} T_\mathrm{c,onset}/\mathrm{d}\Phi \approx \qty{-1.68e-14}{\K\cm\squared}$ is approximately 0.3 times the value predicted by the universal NIEL scaling, yet it remains well within the scatter observed for a broad range of ion-target combinations \cite{SUMM89}. As discussed below, a plausible origin of this deviation is ion channeling.

In the same fluence regime ($\Phi \leq \qty{3.0e15}{\per\cm\squared}$), the normal-state resistivity in the temperature range $\qty{100}{K} \leq T \leq \qty{300}{K}$ exhibits a linear increase characterized by a slope of $\mathrm{d} \rho_\mathrm{N}/\mathrm{d} \Phi \approx \qty{2.9e-19}{\ohm\cm\cubed}$. At higher fluences, $\Phi \geq \qty{4.5e15}{\per\cm\squared}$, superconductivity is fully suppressed, and the YBCO thin film exhibits thermally activated transport at low temperatures.

Comparisons with other literature data on He$^{+}$ irradiation of YBCO must be undertaken with caution. Under conditions of normal (orthogonal) incidence of the ion beam on highly ordered, ultrathin YBCO films, as encountered when employing a HIM, a significant fraction of the ions can undergo channeling along the crystallographic $c$ axis and propagate through the crystal lattice with minimal scattering. In fact, He$^{+}$ ion channeling and backscattering are routinely employed as highly sensitive analytical methods for evaluating the orientational quality and crystallinity of materials \cite{SWAN82R}. In the context of defect engineering, however, such channeling diminishes the effective number of ions that undergo scattering events and are thus available for the generation of point defects.

With the HIM parameters employed in this study, the convergence semi-angle of the ion beam is approximately \ang{0.3}, whereas the Lindhard critical angle for channeling of \qty{2}{\MeV} He$^+$ ions has been reported to be larger, approximately \ang{0.7} \cite{GEER89}. According to Lindhard's classical theory of ion channeling \cite{LIND64}, the critical angle scales with the square root of the inverse ion energy, which implies a broader angular window for channeling of 30-keV ions in YBCO. Furthermore, the observation that $\mathrm{d} T_\mathrm{c,onset}/\mathrm{d} \Phi$ is only 0.3 times the value predicted by a NIEL-based calculation supports the hypothesis that ion channeling plays a significant role in He-FIB irradiations of thin epitaxial YBCO films. Indeed, studies in \qty{50}{\nm} thick YBCO films employing broad-beam 80-keV He$^{+}$ irradiation at \ang{7} inclination, where channeling is absent, reported an approximately threefold enhancement in the irradiation efficiency for $T_c$ suppression \cite{SEFR01} as compared with the present work. These findings also indicate that the reduction in nuclear scattering cross section at higher ion energies has a comparatively minor effect relative to deviations from orthogonal incidence.

%---------------------------------------------------
\subsection{Superconducting transition in magnetic fields}
\label{sec:Hc2}

\begin{figure*}[ht]
\centering
\includegraphics[width=\linewidth]{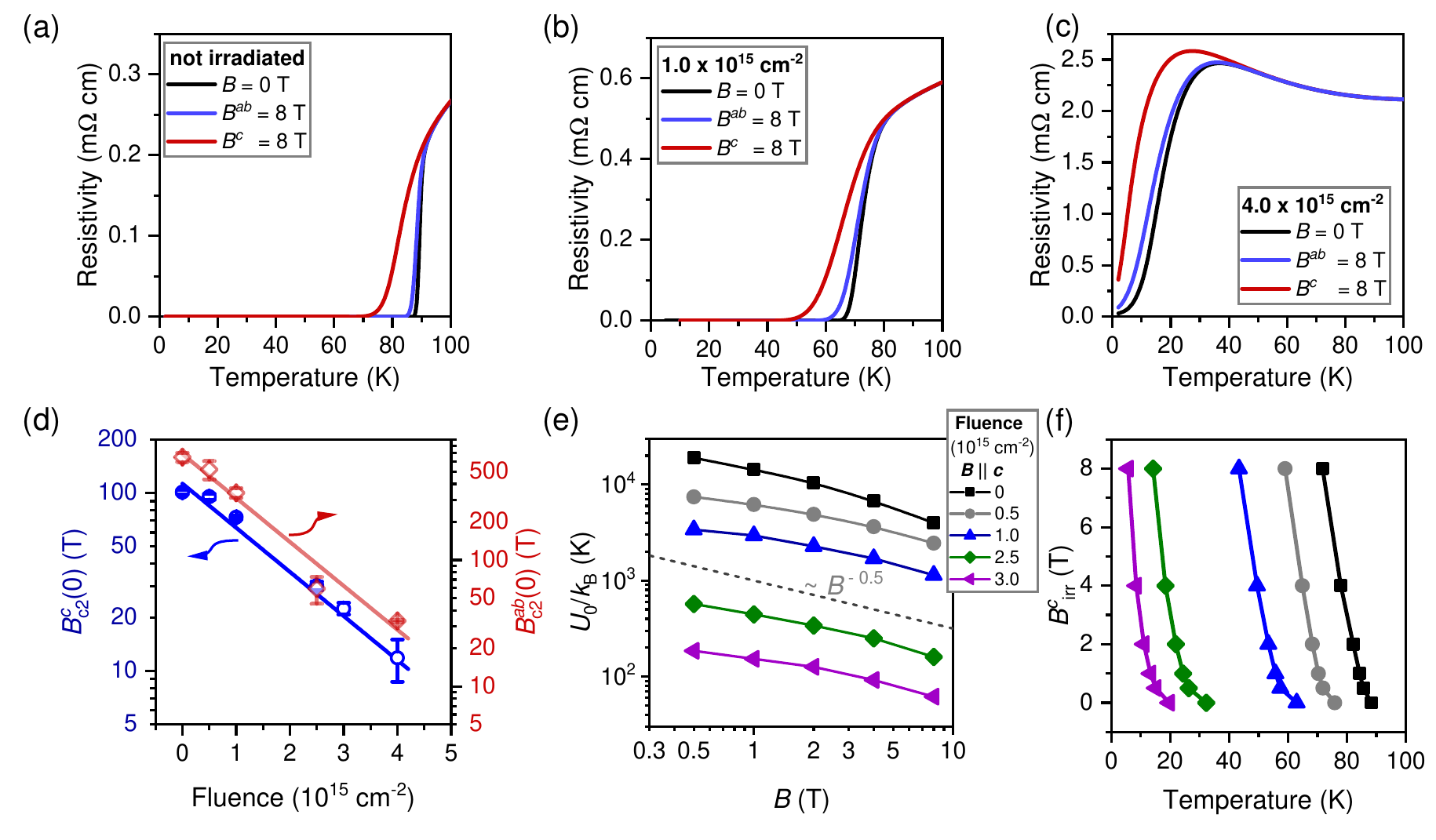}
\caption{(a)-(c) $\rho(T)$ curves at $B=0$ (black lines) and at $B=\qty{8}{\T}$, oriented either as $B^c \parallel c$ (red lines) or $B^{ab}\parallel ab$ (blue lines) for three samples irradiated with $\Phi=0$, 1 and \qty{4.0e15}{cm^{-2}}, respectively. (d) Zero-temperature upper critical fields $B_{\mathrm{c}2}^c$ and $B_{\mathrm{c}2}^{ab}$ for various YBCO bridges irradiated with different 30 keV He$^+$ ion fluences. Error bars denote the 95-\% confidence intervals. The lines represent fits to an exponential decay function. (e) Vortex pinning potential $U_0$ extracted from the slopes of Arrhenius plots of the superconducting transition for $B \parallel c$. The dashed line indicates a power-law dependence $U_0 \sim B^{-0.5}$. (f) Irreversibility lines of various YBCO bridges irradiated with different 30 keV He$^+$ ion fluences. The legend is identical to that in panel (e).}
\label{fig:MR}
\end{figure*}

In HTS, the application of external magnetic fields oriented parallel to the crystallographic $c$ axis leads to a pronounced broadening of the superconducting transition, which is commonly attributed to strong superconducting fluctuations and complex vortex dynamics in the mixed state \cite{RI94}. This field-induced broadening remains clearly discernible following He$^+$ ion irradiation, as demonstrated in Fig.~\ref{fig:MR}(a)-(c). For the purposes of quantitative analysis, we define the critical temperature $T_\mathrm{c}$ as the point of the superconducting transition, where $\rho(T,B) = \rho_\mathrm{onset}(B=0)/2$. Notably, the relative shift $[T_\mathrm{c}(\qty{8}{T})-T_\mathrm{c}(B=0)]/T_\mathrm{c}(B=0)$ increases in magnitude from $-6\,\%$ in the pristine sample to $-37\,\%$ in the microbridge irradiated with a fluence of $\Phi = \qty{4.0e15}{\per\cm\squared}$. This pronounced enhancement of the field-induced suppression of $T_\mathrm{c}$ indicates a significant reduction of the upper critical field $B_{\mathrm{c}2}^c(0)$ by irradiation.

For the subsequent analysis, we define $B_{\mathrm{c}2}^c(T)$ by the values   $T_\mathrm{c}(B)$. The extrapolation to $T=0$ is carried out using the Helfand–Werthamer formalism \cite{HELF66}.
\begin{equation} \label{eq:Bc2}
B_{\mathrm{c}2}^c(0) \simeq - 0.7\cdot T_\mathrm{c}\cdot \frac{\mathrm{d}B_{\mathrm{c}2}^c(T)}{\mathrm{d}T}\bigg|_{T_\mathrm{c}},
\end{equation}
where $\mathrm{d}B_{\mathrm{c}2}/\mathrm{d}T$ is obtained from a linear regression of the data in the magnetic-field interval between \qty{2}{T} and \qty{8}{T}. Within this range of applied fields, the experimental data exhibit a strictly linear dependence, whereas measurements at lower fields display a slight positive curvature and are additionally affected by larger experimental uncertainties. Consequently, these low-field data points were excluded from the fitting procedure.

The extrapolation yields a value of $B_{\mathrm{c}2}^c(0) = \qty{101 \pm 4}{\T}$ for the pristine bridge, which is in good agreement with reports on untwinned 50-nm-thick YBCO samples \cite{WAHL22}. Upon irradiation, $B_{\mathrm{c}2}^c(0)$ decreases rapidly, following an approximately exponential decay, as illustrated in Fig.~\ref{fig:MR}(d). The upper critical field values can additionally be employed to estimate the in-plane coherence length at $T = 0$, defined as $\xi_{ab}(0) = \sqrt{\Phi_0/[2 \pi B_{\mathrm{c}2}^c(0)]}$, where $\Phi_0$ denotes the magnetic flux quantum. The corresponding values of $\xi_{ab}(0)$ are compiled in Table~\ref{tab:properties}.

Since enhanced scattering diminishes the carriers’ normal-state mean-free path $l$, the conventional dirty-limit renormalization of the coherence length, $\xi \rightarrow \sqrt{\xi l}$, would predict a reduction of $\xi_{ab}(0)$ and a concomitant increase of $B_{\mathrm{c}2}^c(0)$ with increasing irradiation fluence. Contrary to this expectation, we detect only a negligible change in the slopes $\mathrm{d}B_{\mathrm{c}2}^c(T)/\mathrm{d}T$ upon irradiation, together with the development of a positive curvature of $B_{\mathrm{c}2}^c(T)$ for $B \leq \qty{1}{T}$. Therefore, the experimentally observed decrease of $B_{\mathrm{c}2}^c(0)$ can be attributed predominantly to the suppression of $T_c$, consistent with Eq.~\eqref{eq:Bc2}. In oxygen-implanted YBCO films, a reduction of $\mathrm{d}B_{\mathrm{c}2}^c(T)/\mathrm{d}T$ has been reported and was ascribed to defect-induced spatial inhomogeneities of $\xi_{ab}(0)$ \cite{ANTO20}. In contrast, our measurements indicate that pronounced spatial modulations of $\xi_{ab}(0)$ are absent in the present samples.

The low-resistivity segment of the superconducting transition can be described by a thermally activated flux-flow behavior, $\rho(B,T) = \rho_0(B) \exp \left[-U_0(B)/(k_\mathrm{B} T)\right]$, where $U_0(B)$ denotes the activation energy associated with vortex motion and $k_\mathrm{B}$ is the Boltzmann constant. From the slopes of the corresponding Arrhenius plots, the quantity $U_0(B)$ is extracted and represented in Fig.~\ref{fig:MR}(e) in terms of the characteristic temperatures $U_0(B)/k_\mathrm{B}$ for different irradiation fluences and magnetic fields. For fluences $\Phi < \qty{3.0e15}{\per\cm\squared}$, the resistivity data obey the Arrhenius relation over approximately three decades in $\rho$. At higher irradiation levels, the low-resistivity tail develops a pronounced positive curvature, which can be attributed to a crossover from purely thermally activated flux motion to a vortex-glass regime \cite{SEFR01}.

Irradiation results in a pronounced reduction of $U_0$ up to a fluence of $\Phi = \qty{3.0e15}{\per\cm\squared}$. For still higher fluences, a reliable determination of $U_0$ is no longer feasible, or superconductivity is completely suppressed. In the pristine film, $U_0$ approximately follows a $B^{-0.5}$ dependence at low magnetic fields; after irradiation, this field dependence becomes less steep at low fields, but tends to recover the $B^{-0.5}$ scaling at higher magnetic fields. Notably, the slopes of the $U_0(B)$ characteristics shown in Fig.~\ref{fig:MR}(e) are very similar for all irradiated bridges.

The introduction of point defects into our twinned YBCO films diminishes the effectiveness of the one-dimensional pinning associated with twin boundaries aligned parallel to the $c$ axis by promoting lateral out-bending and entanglement of the vortex lines. In this resulting entangled vortex state, thermally assisted plastic vortex flow has been analyzed theoretically \cite{VINO90}, yielding a predicted dependence $U_0 \sim B^{-0.5}$. Consistent with this prediction, YBCO films irradiated with 80-keV He$^+$ ions exhibited such a field dependence \cite{SEFR01}. Moreover, in electron-irradiated untwinned YBCO single crystals \cite{FEND95}, as well as in twinned YBCO crystals irradiated with protons \cite{LOPE98}, a scaling behavior $U_0 \sim B^{-0.7}$ was observed and was likewise attributed to enhanced vortex-line wandering that ultimately overcomes the localizing influence of the twin boundaries.

We determine the irreversibility lines $B_{\text{irr}}(T)$ at the ``offset'' temperature $T_{\mathrm{c}0}$, defined by the condition $\rho(T_{\mathrm{c}0}) = 10^{-3}\,\rho(\qty{100}{\K})$ at which the resistivity becomes negligibly small. The results are presented in Fig.~\ref{fig:MR}(f). Notably, the slopes of $B_{\text{irr}}(T)$ exhibit only a minimal change upon irradiation, while the curves are systematically shifted to lower temperatures. This behavior stands in contrast to oxygen-depleted YBCO, where a reduction in the slope has been reported, and the data follow a universal scaling relation $B_{\mathrm{irr}}(T) = B_0^*(1 - T/T_{\mathrm{c}0})^n$ with $n \approx 1.7$ \cite{OSSA92}. In the present samples, the exponent lies in the range $1.5 \leq n \leq 2.8$, evolving from the pristine film to the specimen irradiated at a fluence of $\Phi = \qty{3.0e15}{\per\cm\squared}$.

%---------------------------------------------------
\subsection{Anisotropy}
\label{sec:anisotropy}

The intrinsically layered crystal structure of YBCO gives rise to a pronounced anisotropy in numerous physical properties. When an electrical current is applied parallel to the $ab$ planes and the external magnetic field is rotated within a plane perpendicular to the current direction, the Lorentz force acting on the vortices remains constant. Nevertheless, for the configuration $B \parallel ab$ the vortices become effectively ``locked'' between the CuO$_2$ layers \cite{FEIN90}, which leads to a markedly reduced vortex mobility and, consequently, to a lower resistivity compared with the configuration $B \parallel c$. This behavior is illustrated in Figs.~\ref{fig:MR}(a)–\ref{fig:MR}(c) for an unirradiated reference bridge and two irradiated bridges. Correspondingly, the broadening of the superconducting transition in magnetic fields with $B \parallel ab$ is significantly smaller.

Applying the same procedure as before to estimate $B_{\mathrm{c}2}^{ab}$ for in-plane oriented magnetic fields via Eq.~\eqref{eq:Bc2} yields substantially higher values. For the pristine YBCO film, we obtain $B_{\mathrm{c}2}^{ab} = \qty{652 \pm 56}{T}$, which is comparable to the estimate derived from the slopes of magnetization curves in YBCO single crystals, $B_{\mathrm{c}2}^{ab} = \qty{674}{T}$ \cite{WELP89}. However, such extraordinarily high values may be merely hypothetical, as experiments employing explosive-driven pulsed magnetic fields have demonstrated a transition to the normal state at $B^{ab} = \qty{240}{T}$ at $T = \qty{1.6}{K}$. This discrepancy has been attributed to the paramagnetic (Pauli) limiting effect on superconductivity, arising from the Zeeman splitting in extremely high magnetic fields \cite{DZUR98}. Nonetheless, our estimate of the out-of-plane coherence length at $T=0$, $\xi_c(0) = \qty{0.28}{nm}$, obtained from the Ginzburg–Landau relation $\xi_{c}(0) = \Phi_0/[2 \pi B_{c2}^{ab}(0) \xi_{ab}(0)]$, remains valid.

Analogous to $B_{\mathrm{c}2}^c(0)$, the in-plane upper critical field $B_{\mathrm{c}2}^{ab}(0)$ also exhibits a rapid, approximately exponential decrease with increasing irradiation fluence, as shown in Fig.~\ref{fig:MR}(d). From these quantities, the superconducting anisotropy can be determined as $\gamma = B_{\mathrm{c}2}^{ab}(0)/B_{\mathrm{c}2}^c(0) = \xi_{ab}(0)/\xi_c(0)$. For the pristine film, we obtain $\gamma = 6.5$, which is a representative value for YBCO \cite{FARR90}. Upon irradiation, the anisotropy is significantly reduced, as summarized in Table~\ref{tab:properties}. This behavior contrasts with observations in oxygen-deficient YBCO single crystals \cite{CHIE94} and thin films \cite{GOB95a}, where the anisotropy increases as $T_\mathrm{c}$ is suppressed. The opposite trend emphasizes that distinct microscopic mechanisms govern the suppression of superconductivity in the two cases.

In oxygen-depleted samples, the removal of oxygen from the O(1) chain sites weakens the interlayer coupling between the CuO$_2$ planes. By contrast, irradiation predominantly generates Frenkel pairs involving oxygen atoms, which interact with the layered crystal structure of YBCO in a different manner. Furthermore, we propose that the observed reduction in anisotropy is consistent with a scenario in which a fraction of the displaced oxygen atoms occupies interstitial positions, and not exclusively the O(5) sites, particularly at higher fluences. This interpretation is compatible with molecular dynamics simulations that predict a finite probability for atoms to be relocated to interstitial sites \cite{GRAY22}.

%---------------------------------------------------
\subsection{Hall effect}
\label{sec:Hall}

\begin{figure*}[ht]
\centering
\includegraphics[width=\linewidth]{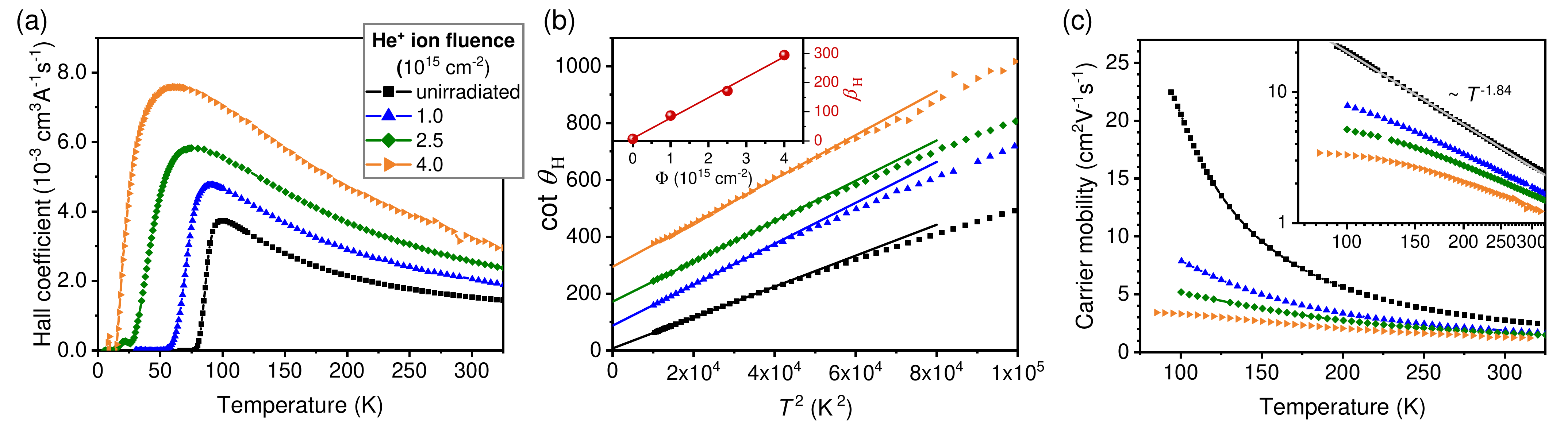}
\caption{(a) Hall coefficient $R_\mathrm{H}$ of a YBCO thin film as a function of temperature before and after irradiation with 30~keV He$^+$ ions. (b) Cotangent of the Hall angle at $B=\qty{1}{\T}$ as a function of $T^2$. The solid lines represent fits to the data in the temperature range $\qty{100}{\K} \leq T \leq \qty{200}{\K}$. Their slopes are associated with the carrier density, whereas the intercepts are indicative of the concentration of scattering centers. Inset: Intercepts $\beta_\mathrm{H}$ obtained from linear fits to the $\cot \theta_\mathrm{H}(T^2)$ curves as a function of ion fluence $\Phi$. The solid lines represents a linear fit. (c) Normal-state carrier mobility of a YBCO thin film as a function of temperature before and after irradiation with 30~keV He$^+$ ions. Inset: Same data in a log-log representation, demonstrating a power-law dependence for the unirradiated bridge.}
\label{fig:Hall}
\end{figure*}

Hall-effect measurements, when analyzed in conjunction with resistivity data, yield a more complete characterization of the underlying charge-transport mechanisms. In HTS, however, the Hall effect displays an anomalous temperature dependence that precludes the straightforward application of the Drude model to determine the carrier concentration using $n_\mathrm{H} = (R_\mathrm{H} q)^{-1}$, where $q$ denotes the elementary charge of electrons or holes, respectively, and $R_\mathrm{H}$ the Hall coefficient. In particular, for YBCO, as for most HTS, $n_\mathrm{H}$ increases approximately linearly with the temperature $T$ in the normal state \cite{WUYT93}.

The Hall coefficient of the YBCO thin films, determined at $B^c = \qty{8}{\T}$ before and after ion irradiation with moderate fluence, is shown in Fig.~\ref{fig:Hall}(a). The pristine specimen exhibits the characteristic, but unconventional, Hall response expected for nearly optimally doped YBCO. Following our aforementioned interpretation of the resistivity data, ion irradiation is found to leave the mobile carrier concentration essentially unchanged. Therefore, one might expect $n_\mathrm{H}$ and, hence, $R_\mathrm{H}$ be unaffected by irradiation. However, $R_\mathrm{H}$ exhibits a substantial enhancement upon irradiation and continues to display a pronounced dependence on temperature.

It is widely recognized that an analysis of the Hall angle, $\tan \theta_\mathrm{H} = R_\mathrm{H} B/ \rho = \mu B$, which is directly related to the carrier mobility $\mu$, provides deeper physical insight and reveals a higher degree of universality across different doping levels in HTS \cite{BARI22}. The Hall angle is commonly described by the semiempirical relation originally proposed by Anderson \cite{ANDE91},
\begin{equation}
\label{eq:Anderson}
\cot \theta_\mathrm{H} = \alpha_\mathrm{H} T^2 + \beta_\mathrm{H},
\end{equation}
where $\alpha_\mathrm{H}$ and $\beta_\mathrm{H}$ are assumed to be temperature-independent parameters. The experimental Hall-effect data, displayed in Fig.~\ref{fig:Hall}(b), are analyzed within this phenomenological framework. In this representation, a larger slope $\alpha_\mathrm{H}$ is associated with a higher charge-carrier concentration \cite{WUYT93}, whereas $\beta_\mathrm{H}$ increases linearly with the density of pair-breaking defects \cite{CHIE91a}. Within this context, it is apparent that irradiation only weakly modifies $\alpha_\mathrm{H}$, while the intercept $\beta_\mathrm{H}$ increases approximately linearly with irradiation fluence, as demonstrated in the inset of Fig.~\ref{fig:Hall}(b). In stark contrast to our observations, a reduction of the oxygen content in YBCO leads to a decrease in the density of mobile carriers and, consequently, to a significantly smaller slope $\alpha_\mathrm{H}$ \cite{WUYT93}. The Hall-effect measurements, therefore, provide additional support for our earlier conclusion that He$^+$ irradiation reduces $T_\mathrm{c}$ primarily by creating defects.

The temperature dependence of the carrier mobility in irradiated YBCO is presented in Fig.~\ref{fig:Hall}(c). The inset demonstrates that the mobility of the pristine film follows a power-law dependence, $\mu \sim T^{-1.84}$, in the normal state over the entire investigated temperature range. This exponent is close to the $T^{-2}$ behavior expected for Fermi-liquid (hole–hole) scattering and is characteristic of HTS. Following irradiation, the mobility $\mu(T)$ decreases monotonically at any point in the measured temperature range. Moreover, the temperature dependence of $\mu(T)$ is reduced, which can be primarily attributed to the enhanced contribution of defect scattering, as reflected by the increasing intercept in Fig.~\ref{fig:Hall}(b). An analogous correlation between defect concentration and Hall mobility has been reported for YBa$_2$Cu$_{3-x}$Zn$_x$O$_{7-\delta}$ single crystals, in which defects were systematically introduced by substituting Cu with isovalent Zn atoms \cite{CHIE91a}.

As has been noted previously \cite{VALL89}, the carrier density
does not vary sufficiently under moderate ion irradiation to account for the pronounced suppression of $T_c$, whereas the enhanced carrier scattering is manifested by a strong reduction in mobility. It is therefore instructive to estimate the average carrier mean-free path $l$ and compare it with the in-plane coherence length $\xi_{ab}(0)$. As a first-order approximation, we employ the single parabolic band expression $l = \left( 3 \pi^2 n \right)^{1/3} \mu \hbar/e$, with a carrier density $n = \qty{2e22}{\per\cm\cubed}$, corresponding to 1.16 holes per Cu atom and assumed to remain unchanged under irradiation, consistent with our previously discussed findings. The mobility $\mu$ is evaluated at 100\,K. The resulting mean-free-path values are listed in Table~\ref{tab:properties}. 

For the pristine sample, the condition $l > \xi_{ab}(0)$ indicates that the system is in the clean limit. Ion irradiation, however, decreases $l$ while simultaneously increasing $\xi_{ab}(0)$, thereby driving a crossover to the dirty limit at a fluence of $\Phi \simeq \qty{2.5e15}{\per\cm\squared}$. This crossover should be taken into account in any quantitative analysis of the physical properties of ion-irradiated YBCO.

%---------------------------------------------------

\begin{table*}
\begin{center}
\caption{Magnetotransport characteristics of YBCO thin-film microbridges (sample A) prior to and following irradiation with 30~keV He$^+$ ions. Data from sample B are marked by an asterisk. The corresponding physical quantities are defined in detail in the main text and are listed with their 95\% confidence ranges. Values in brackets may be influenced by constraints in the available data range.}
\begin{tabular}{|c|c|c|c|c|c|c|c|c|c|} \hline
Fluence & $T_{c}$ & $T_{c-onset}$\ & $T_{c0}$ & $B_{c2}^c(0)$ & $B_{c2}^{ab}(0)$ & $\xi_{ab}(0)$ & $\xi_{c}(0)$ & $\gamma$ & $l(\qty{100}{\K})$\\
(\SI{}{\per\centi\meter\squared}) & (K) & (K) & (K) & (T) & (T) & (nm) & (nm) & & (nm) \\ \hline\hline
unirradiated & 88.8 & 90.1 & 86.2 & $101.1 \pm 4$ & $651.8 \pm 56$ & $1.80 \pm 0.03$ & $0.28 \pm 0.03$ & $6.5 \pm 0.6$ & [...] \\ 
             & 90.0$^*$ & 91.4$^*$ & 88.3$^*$ & $110.8 \pm 8^*$ & $(637.3)^*$ & $1.72 \pm 0.06^*$ & $(0.30)^*$ & $5.8 \pm 1.8^*$ & 11$^*$  \\ \hline
\num{0.5e15} & 80.6 & 84.0 & 75.9 & $95.3 \pm 4$ & $520.5 \pm 87$ & $1.90 \pm 0.04$ & $0.34 \pm 0.06$ & $5.5 \pm 1.0$ & 5.6 \\ \hline
\num{1.0e15} & 70.7 & 76.8 & 63.0 & $73.0 \pm 2$ & $340.8 \pm 31$ & $2.12 \pm 0.03$ & $0.45 \pm 0.05$ & $4.7 \pm 0.5$ & 4.4 \\ \hline
\num{1.5e15} & (59.5) & (66.1) & 49.0 & [...] & [...] & [...] & [...] & [...] & [...]\\ \hline
\num{2.5e15} & 43.7 & 51.1 & 32.3 & $30.1 \pm 2$ & $59.3 \pm 15$ & $3.31 \pm 0.10$ & $ 1.68 \pm 0.41$ & $2.0 \pm 0.5$ & 2.9 \\ \hline
\num{3.0e15} & 33.8 & 42.0 & 19.7 & $22.4 \pm 2$ & [...] & $3.83 \pm 0.15$ & [...] & [...] & [...] \\ \hline

%\num{3.5e15} & 12.7 & 23.4 & < 2 & $8.9 \pm 2$ & - & $6.07 \pm 0.69$ & - & - & - \\ \hline
\num{4.0e15} &  10.8 & 19.0 & < 1.8 & $11.9 \pm 4$ & $33.0 \pm 1$ & $5.27 \pm 0.70$ & $1.89 \pm 0.26$ & $2.8 \pm 0.8$ & 1.8 \\ \hline
\end{tabular}
\label{tab:properties}
\end{center}
\end{table*}

%---------------------------------------------------
\section{Summary and Conclusions}

We have investigated the evolution of structural and electronic properties of epitaxial YBCO thin films on LSAT subjected to 30-keV He$^+$ irradiation over a wide fluence range, from pristine material to $\Phi=\qty{1e16}{\cm^{-2}}$, where superconductivity is fully suppressed and the crystalline framework is largely destroyed. By combining x-ray diffraction, Raman scattering, and comprehensive magnetotransport measurements, we establish a consistent picture in which irradiation predominantly introduces disorder in the form of oxygen-related Frenkel defects rather than driving a significant change of oxygen stoichiometry or mobile carrier concentration.

XRD reveals a progressive degradation of crystallinity with fluence: the (00$l$) peaks broaden, shift to smaller angles, and lose intensity markedly for $\Phi \gtrsim \qty{5e15}{\cm^{-2}}$, consistent with the onset of strong disorder and amorphization. The out-of-plane lattice parameter $c$ increases approximately linearly with fluence and reaches values exceeding those typical of oxygen-depleted YBCO. The lattice parameter $a$ remains essentially constant while $b$ decreases with fluence, resulting in an orthorhombic-to-tetragonal transition at $\Phi \ge \qty{8e15}{\cm^{-2}}$.

Raman spectroscopy supports this conclusion by the absence of hallmark signatures of oxygen depletion, most notably the lack of the \qty{228}{\per\cm} mode associated with broken CuO chains. The Cu(2) mode near \qty{144}{\per\cm} shifts only weakly with fluence, while its spectral weight increases, consistent with a growing tetragonal fraction without substantial loss of oxygen. Together with the systematic rise of the Raman background, the spectra indicate increasing structural disorder that is compatible with the displacement of O(1) atoms into O(5) sites and other interstitial positions.

Electrical transport measurements quantify how this defect landscape tunes superconductivity and normal-state conduction. Irradiation increases the normal-state resistivity and suppresses $T_\mathrm{c}$ and the carrier mobility, while the invariance of the slopes of resistivity and inverse Hall angle implies an approximately fluence-independent carrier concentration. In contrast, the residual terms of both quantities increase strongly, evidencing enhanced defect scattering. The suppression of the onset temperature of superconductivity is well described by Abrikosov--Gor'kov pair breaking using a single fit parameter, enabling practical predictions of $T_\mathrm{c}$ suppression from experimental fluence and simulated dpa values.

Magnetotransport further demonstrates that irradiation not only reduces $T_\mathrm{c}$ but also renormalizes key mixed-state parameters. The upper critical fields extrapolated to $T=0$ decrease approximately exponentially with fluence for both $B \parallel c$ and $B \parallel ab$, with the reduction of $B_{c2}^c(0)$ largely attributable to the suppression of $T_\mathrm{c}$ rather than a strong change of $\mathrm{d}B_{c2}/\mathrm{d}T|_{T_c}$. The vortex activation energy $U_0$ decreases substantially with increasing fluence, consistent with weakened effective pinning in these twinned films and enhanced vortex wandering; irreversibility lines shift to lower temperatures while their slopes remain comparatively robust. Significantly, the anisotropy $\gamma=B_{c2}^{ab}(0)/B_{c2}^c(0)$ is reduced from its pristine value ($\gamma \approx 6.5$), opposite to the trend commonly observed in oxygen-depleted YBCO. This provides an additional, independent signature that He$^+$ irradiation suppresses superconductivity via disorder (pair breaking and altered interlayer coherence) rather than through reduced hole doping. A crude mean-free-path estimate indicates a crossover from the clean to the dirty limit at high fluence (around $\Phi=\qty{2.5e15}{\cm^{-2}}$), which is relevant for modeling irradiated regions and for interpreting device performance.

In conclusion, 30-keV He$^+$ irradiation provides a controlled route to engineer superconducting and normal-state properties of YBCO thin films predominantly through disorder generation. For device patterning with helium ion microscopy, these results delineate two practically significant fluence scales: (i) a functional tuning regime up to $\Phi \sim \qty{4e15}{\cm^{-2}}$, where $T_\mathrm{c}$, resistivity, and vortex dynamics are continuously adjustable while crystallinity remains largely intact, and (ii) complete suppression of superconductivity at $\Phi \gtrsim \qty{4.5e15}{\cm^{-2}}$ that enables the formation of robust barriers for, e.g., Josephson junctions. At substantially higher fluence $\Phi \gtrsim \qty{8e15}{\cm^{-2}}$, dominant amorphization is expected to compromise spatial resolution and is therefore unfavorable for nanoscale circuit fabrication. These quantitative structure--property correlations provide a foundation for predictive defect-engineering protocols and for integrating YBCO into irradiation-defined superconducting electronics. It should be emphasized that the present study concerns large-area irradiation. Irradiation confined to nanoscale regions \cite{CART26} introduces additional complexity, leading to pronounced size effects arising from the finite lateral dimensions of the irradiated areas embedded within the surrounding non-irradiated film.

%---------------------------------------------------
\section*{Acknowledgements}

P.\,A.\,K thanks D. Baurecht for support with the Raman measurements at the Faculty Center for Nanostructure Research at the University of Vienna. Parts of this research were carried out at IBC at the Helmholtz-Zentrum Dresden-Rossendorf e.~V., a member of the Helmholtz Association. B.\,A. acknowledges support from Aktion \"Osterreich-Slowakei, A\"OSK-Initiativprojektförderung der Aktion (Grant No. 2025-03-15-004), the Federal Ministery of Women, Science and Research of the Republic of Austria and the Ministry of Education, Research, Development and Youth of the Slovak Republic. This research was funded in whole, or in part, by the Austrian Science Fund (FWF) Grant-DOI: 10.55776/I4865 and by the German Research Foundation (DFG), grants KO~1303/16-1 and Go~1106/7-1. For the purpose of open access, the authors have applied a CC BY public copyright license to any Author Accepted Manuscript version arising from this submission. The publication is based upon work from COST Actions CA21144 (SuperQuMap), CA19140 (FIT4NANO), and CA23134 (Polytopo) supported by COST (European Cooperation in Science and Technology). 

\bibliography{irrad}

\end{document}